\documentclass[11pt,twoside]{article}
\usepackage{asp2010}

\resetcounters

\begin{document}

\title{Astro-WISE Information System}
\author{Willem-Jan Vriend, Edwin A. Valentijn, Andrey Belikov, Gijs A. Verdoes Kleijn
\affil{Kapteyn Astronomical Institute, P.O. Box 800, 9700 AV Groningen, the Netherlands}
}

\begin{abstract}
Astro-WISE is a scientific information system for the data processing of optical images. In this paper we review main features of Astro-WISE and describe the current status of the system. 
\end{abstract}

\section{Introduction}

Astro-WISE~\footnote{http://www.astro-wise.org}~\citep{Astro-WISE} stands for Astronomical Wide-field Imaging System for Europe. The system was initially developed to support
the data processing of the Kilo Degree Survey (KiDS)~\citep{P157_adassxxi} on
the VLT (Very Large Telescope) Survey Telescope (VST\footnote{http://www.eso.org/public/teles-instr/surveytelescopes/vst.html}). 
The VST is a 2.61 m diameter imaging telescope installed on the Paranal Observatory in Chile. The 
instrument installed on the telescope is OmegaCAM, a large format (16k x 16k) CCD camera which will produce up to 
15 TB of raw images per year. This amount is multiplied at least by a factor of 3 by the data processing.

Astro-WISE was planned as a storage and data processing system for KiDS, but, with time,
grew up to a general astronomical information system. It has developed into a much wider data processing information system
which can be used in many other disciplines. The idea behind Astro-WISE is to have the data model, data and data processing in a single system.

At the same time, such a system should be shared by a number of institutes and sites, as the scientific community working 
with the data in the system is widely distributed and each group is coming to the system with their own resources 
(data storage and computing facilities). Moreover, each group is developing a part of the software implemented in the 
system. The users must be able to form communities (groups) to process the data for the same instrument or project.

\section{Basic Principles of the System}

The development of the Astro-WISE information system started from the very practical challenge:  enable a community of researchers distributed over the world to process
the data of the OmegaCAM 256 Megapixel camera imaging survey. These scientists should be able to evaluate the quality of the data, apply a number of calibrations, share the data in the team and employ
distributed resources of PB scale of data storage and Tflops capacity in data processing.

From this our basic requirements on the system are derived:
\begin{itemize}
\item[*] Scalability of the system: any part of the system, i.e. data storage, data processing, metadata management, should be scalable with the increase of incoming data and a number of users 
involved in the data processing. The system should be scalable to the data processing algorithms and pipelines allowing to implement new pipelines and reprocess the same data with new algorithms. 
The scalability to the data mining should be implemented, i.e., the system should provide all possible ranges of requests from the retrieval of a single data item by identifier to a complicated 
archive study involving multiple complex queries.
\item[*] Distributed system: the system allows any activity to be distributed among different users and different sites where the system is implemented.
\item[*] Traceability: all activity in the system should leave a clear footprint so that it will be possible to trace the origin of any changes in the data and find an algorithm, program and user 
who created a data item.
\item[*] Adaptability. The system allows for a number of different scientific use-cases and provides resources, pipelines and expertise to perform data processing according to the user's interests.
\end{itemize}

Requirements were set on the common data model realized in the system. The common data model is the core of Astro-WISE and implements the following features:
\begin{itemize}
\item[1.] {\bf Inheritance of data objects}.  Using object oriented programming, all objects within the system can inherit key properties of the parent object, all these properties are made persistent (i.e., stored in a database).
\item[2.] {\bf Full lineage}. The linking (associations or references, or joins) between object instances in the database is maintained completely. Each data item in the system can be traced back to its origin. The tracing of the data object can be both forward and backward, for example, this makes it possible to find which raw frames were used to determine magnitudes, shapes and position for this particular source and, at the same time, which sources were extracted on the particular raw frame.
\item[3.] {\bf Consistency}. At each processing step, all processing parameters and the inputs which are used, are kept within the system. Astro-WISE can keep the previous versions of all data items along with all parameters used to produce them and all dependencies between objects.
\item[4.] Embarrassingly {\bf parallel and distributed processing}, the administration of asynchronous processing is naturally recorded in the metadata layer
\end{itemize}

The programming of both Astro-WISE pipelines and also programs employs:
\begin{itemize}
\item[1.] {\bf Component based software engineering (CBSE)}. This is a modular approach to software development, each module can be developed independently and wrapped in the base language of the system (Python) to form a pipeline or workflow.
\item[2.] {\bf An object-oriented common data model used throughout the system}. This means that each module, application and pipeline will deal with the unified data model for the whole cycle of the data processing from the raw data to the final data product.
\item[3.] {\bf Persistence of all the data model objects}. Each data product in the data processing chain is described as an object of a certain class and saved in the archive of the specific project along with
the parameters used for data processing.
\end{itemize}

\section{Realization and Current Status}

\begin{figure*}
\centering
\includegraphics[width=0.80\textwidth]{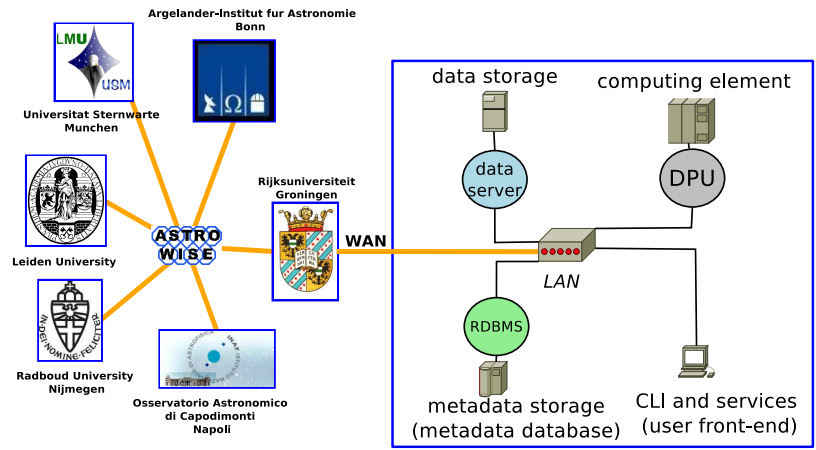}
\caption{Astro-WISE architecture. The Groningen node is shown. The node consistst of a metadata database (Oracle RAC 11g), computing element (DPU submitting jobs to HPC in Donald Smits Center for 
Information Technology), data storage (Astro-WISE dataservers) and a number of web services.}
\label{fig:AW}
\end{figure*}

Astro-WISE is a federated system distributed over Europe. Currently nodes of Astro-WISE are installed in the Groningen University, Radboud University Neijmegen, Leiden University (The Netherlands),
Bonn Argelander-Institut f\"ur Astronomy, Universit\"ats-Sternwarte M\"unchen (Germany) and the Observatorio Astronomico di Capodimonte (Italy). Each data item in Astro-WISE is stored in a file
with an unique filename registered in metadata database. Astro-WISE allows users to share data in projects forming groups responsible for the processing of data for a particular instrument or survey.
For example, Astro-WISE is used for KiDS, the Coma Legacy Survey\footnote{http://www.astro-wise.org/projects/COMALS/} and various WFI surveys.

An Astro-WISE node in full deployment consists of a metadata database, computing element, data storage element (dataservers), web services and an Astro-WISE user environment. The data processing 
can be submitted to one of the Astro-WISE computing elements or BiG Grid processing element via the Distributed Processing Unit (DPU).

Each data item in Astro-WISE is an object of a corresponding class, for example, all reduced images are objects of the {\it ReducedScienceFrame} class and keep links to the scientific images as well as
the calibration images used for the data processing, processing parameters and quality control parameters. This information is stored in the metadata part of the data item and available to the user according 
to the user's access rights. It is possible to use a number of Astro-WISE web services to process the data or to verify the quality of the processed data~\citep{D5_adassxxi}. 
In addition to web services the user can write a Python script to build his own pipeline from a library of pre-defined methods and classes for the data processing. The processed data will be stored 
as private data of the user till the user decides to share it within the project or to publish it in Astro-WISE or Virtual Observatory.  

The Astro-WISE system supports data processing from raw images not only for OmegaCAM but for a wide range of instruments, including WFC, WFI, Megacam, SuprimeCAM, Large Binocular Camera and 
others~\footnote{http://www.astro-wise.org/portal/instruments\_index.shtml}. Astro-WISE provides users with a number of catalogs and surveys, including USN0-B1, 2MASS PSC, SDSS DR7, UKIDSS DR3 and others,
which can be combined with the data produced by the user in a new data product.

In 2011 Astro-WISE has a storage capacity of 1.6 PB of data and processing capacity of 10 Tflops distributed over Astro-WISE partners. In addition users of Astro-WISE can employ processing 
elements of BiGGrid\footnote{http://www.biggrid.nl}. 

In the recent years in addition to the data processing of the KiDS survey the Astro-WISE concept was used to develop information system sfor LOFAR Long-Term Archive~\citep{P009_adassxxi} and
Multi Unit Spectroscopic Explorer data processing system~\citep{P117_adassxxi}.

\acknowledgements Astro-WISE is an on-going project which started from a FP5 RTD programme funded
by the EC Action ``Enhancing Access to Research Infrastructures''. This work is
supported by FP7 specific programme ``Capacities - Optimising the use and
development of research infrastructures''.

\bibliography{P160}

\end{document}